\RequirePackage{ifpdf}
\ifpdf % We are running pdfTeX in pdf mode
\documentclass[pdftex]{sigma}
\else
\documentclass{sigma}
\fi

\begin{document}

\renewcommand{\PaperNumber}{110}

\FirstPageHeading

\renewcommand{\thefootnote}{$\star$}

\ShortArticleName{Alternative Method for Determining the Feynman Propagator}

\ArticleName{Alternative Method for Determining\\ the Feynman
Propagator of a Non-Relativistic\\ Quantum Mechanical Problem\footnote{This paper is a contribution to the Proceedings
of the Seventh International Conference ``Symmetry in Nonlinear
Mathematical Physics'' (June 24--30, 2007, Kyiv, Ukraine). The
full collection is available at
\href{http://www.emis.de/journals/SIGMA/symmetry2007.html}{http://www.emis.de/journals/SIGMA/symmetry2007.html}}}

\Author{Marcos MOSHINSKY~$^\dag$, Emerson SADURNI~$^\dag$ and Adolfo DEL CAMPO~$^\ddag$}

\AuthorNameForHeading{M. Moshinsky, E. Sadurn\'i and A. del Campo}

\Address{$^\dag$~Instituto de F\'{\i}sica
 Universidad Nacional Aut\'onoma de M\'exico, \\
$\phantom{^\dag}$~Apartado Postal  20-364,
 01000 M\'exico D.F., M\'exico}
\EmailD{\href{mailto:moshi@fisica.unam.mx}{moshi@fisica.unam.mx}, \href{mailto:sadurni@fisica.unam.mx}{sadurni@fisica.unam.mx}}

\Address{$^\ddag$~Departamento de Qu\'imica-F\'isica, Universidad del
Pa\'is Vasco, Apdo. 644, Bilbao, Spain}
\EmailD{\href{mailto:qfbdeeca@lg.ehu.es}{qfbdeeca@lg.ehu.es}}

\ArticleDates{Received August 21, 2007, in f\/inal form November
13, 2007; Published online November 22, 2007}

\Abstract{A direct procedure for determining the propagator
associated with a
 quantum mechanical problem was given by the Path
Integration Procedure of Feynman. The Green function, which is the
Fourier Transform with respect to the time variable of the
propagator, can be derived later. In our approach, with the help of
a Laplace transform, a direct way to get the energy dependent Green
function is presented, and the propagator can be obtained later with
an inverse Laplace transform. The method is illustrated through
simple one dimensional examples and for time independent potentials,
though it can be generalized to the derivation of more complicated
propagators.}


\Classification{81V35; 81Q05}

\section{Introduction}

It is well know that quantum mechanics acquired its f\/inal
formulation in 1925--1926 through fundamental papers of Schr\"odinger
and Heisenberg. Originally these papers appeared as two independent
views of the structure of quantum mechanics, but in 1927
Schr\"odinger established their equivalence, and since then one or
the other of the papers mentioned have been used to analyze quantum
mechanical systems, depending on which method gave the most
convenient way of solving the problem. Thus the existence of
alternative procedures to solve a given problem can be quite
fruitful in deriving solutions of it.

In the 1940's Richard Feynman, and later many others, derived a
propagator for quantum mechanical problems through a path
integration procedure.  In contrast with the Hamiltonian emphasis in
the original formulation of quantum mechanics, Feynmans approach
could be referred to as Lagrangian and it emphasized the propagator
$K(x,t,x', t')$ which takes the wave function $\psi (x', t')$ at the
point $x'$ and time $t'$  to the point $x,$ at time $t$, i.e.
                \begin{gather}
              \psi (x,t) = \int K (x,t,x',t') \psi (x', t')dx'.
                           \label{mf1}
        \end{gather}

While this propagator could be derived by the standard methods of
quantum mechanics, Feynman invented a procedure by summing all time
dependent paths connecting points $x$, $x'$ and this became an
alternative formulation of quantum mechanics whose results coincided
with the older version when all of them where applicable, but also
became relevant for problems that the original methods could not
solve.   Feynmans procedure f\/irst led to the propagator $K(x, t, x',
t')$ and then by a Laplace transform to the corresponding Green
function $G(x,x',E)$ with $E$ being the energy. We found Feynmans
method for deriving the propagator, though entirely correct,
somewhat cumbersome to use, and thus tried to look for alternative
procedures.

As we mentioned before, in Feynmans approach the f\/irst step is
deriving the propagator $K(x,t,x', t')$ and later the energy
dependent Green functions $G(x,x', E)$. In this paper we invert the
procedure, we start by deriving the $G(x,x', E)$ which is a simpler
problem, at least in the one dimensional single particle case we
will be discussing here. Once we have $G(x,x', E)$ the $K(x,t,x',
t')$ is given by the inverse Laplace transform and can be written as
                      \begin{gather}
           K (x,x',t) = \frac{1}{2\pi\hbar i}
           \int^{i\hbar c+\infty}_{i\hbar c -\infty}
           \exp(-iE t/\hbar) G (x,x',E) dE,
                           \label{mf2}
        \end{gather}
 where $c$ is a constant that allows the upper line $i \hbar c +E$
 in the complex plane of $E$ to be above all the poles of $G(x,x',
 E)$.
 For compactness in the notation from now on we will take $t'=0$ so
 we write $K(x,t,x', t')$ as $K(x,x', t)$.

The real hard part in our approach will be the determination by \eqref{mf2}
of $K(x,x', t)$ but this is a well def\/ined problem in mathematics
and procedures have been developed to solve it.

This is then the program we plan to follow. In Section~\ref{section2} we show
that for a single particle in one dimension (the initial case of our
analysis) all we need to know are two independent solutions
$u^\pm_E$ of the equation.
\[
           \left[\frac{-\hbar^2}{2m} \frac{d^2}{dx^2} + V(x) - E\right] u^\pm_E
           (x,E) = 0
                           %\label{mf3}
\]
to be able to derive $G(x,x', E)$ in Section~\ref{section3}.

We then consider in Section~\ref{section4}, three elementary cases, the free one
dimensional particle, the corresponding one with a $\delta$
interaction at the origin $x=0$, and the harmonic oscillator. In the
f\/irst two cases the integral \eqref{mf2} is trivial to evaluate. In the case
of the harmonic oscillator the evaluation of \eqref{mf2} requires a more
careful analysis but it can be carried out. In all three cases our
f\/inal result is identical to the one presented in the book of Grosche and
Steiner~\cite{1} that use Feynmans method to derive the results.

Thus we have an alternative method for deriving $K(x,x', t)$ but it
remains to be shown that it can be applied to more particles in more
dimensions and of arbitrary angular momenta and whether the analysis
can be extended to relativistic as well as time dependent problems.

What ever results may be obtained
in the future, it seems that our alternative approach follows more closely the standard
procedures of quantum mechanics and could be useful in a~simpler derivation of the
propagators.

\section{The Hamiltonian of the problem and the equation\\ for the
propagator}\label{section2}

We start with the simplest Hamiltonian of one particle in one
dimension, i.e.
\[
        H=-\frac{\hbar^2}{2m} \frac{d^2}{dx^2} + V(x)
                          % \label{mf4}
\]
with thus far an arbitrary potential $V(x)$.

From the equation (\ref{mf1}) that def\/ines the properties of the
propagator it must satisfy the equation
                      \begin{gather}
    \bigg[-\frac{\hbar^2}{2m} \frac{\partial^2}{\partial x^2} + V(x)
              - i\hbar \frac{\partial}{\partial t}\bigg]
              K (x,x',t)= 0
                           \label{mf5}
        \end{gather}
and besides if $t=0$ it becomes
                          \begin{gather}
              K (x,x',0) = \delta (x-x').
                           \label{mf6}
        \end{gather}

We proceed now to take the Laplace transform of (\ref{mf5})
\begin{gather*}
            \int^\infty_0 \exp (-st) \left[ - \frac{\hbar^2}{2m}
            \frac{\partial^2}{\partial x^2} + V(x) - i\hbar \frac{\partial}{\partial t}
                \right] K(x,x',t)dt \nonumber\\
 \qquad {}= -  \frac{\hbar^2}{2m} \frac{\partial^2 \bar G
                (x,x',s)}{\partial x^2} + V(x) \bar G(x,x',s) - i\hbar
                \int^\infty_0 \exp (-st) \frac{\partial K(x,x',t')}{\partial t} dt=0,
                                           %\label{mf7}
\end{gather*}
where
                  \begin{gather}
              \bar G(x,x',s) \equiv \int^\infty_0 e^{-st} K (x, x',
              t)dt.
                           \label{mf8}
\end{gather}
We note though that
\begin{gather}
        \int^\infty_0 \exp(-st) \frac{\partial K(x,x',t)}
                            {\partial t} dt = \int^\infty_0
        \frac{\partial}{\partial t} \big[ e^{-st} K(x,x',t)\big]
        dt + s\int^\infty_0 e^{-st} K(x,x',t) dt\nonumber\\
\qquad{}  = -\delta (x-x') + s \bar G(x,x',s),
      \label{mf9}
\end{gather}
where we made use of \eqref{mf6} and \eqref{mf8}.

With the help of \eqref{mf9} we see that $\bar G(x,x',s)$ satisf\/ies
            \begin{gather}
    \left[ - \frac{\hbar^2}{2m} \frac{d^2}{dx^2} + V(x)
    - i\hbar s\right] \bar G (x,x',s) = -i \hbar \delta (x-x'),
                           \label{mf10}
        \end{gather}
where we now have that the partial derivative with respect to $x$
becomes the ordinary one as there is no longer a time variable.

We integrate~(\ref{mf10}) with respect to~$x$ in the interval
$x'-\epsilon \leq x \leq x' + \epsilon$ and in the limit $\epsilon
\to 0$ obtain two equations
\begin{gather}
    \left[-\frac{\hbar^2}{2m} \left(\frac{d\bar G}{dx}\right)_{x=x'+0}
    + \frac{\hbar^2}{2m}
    \left(\frac{d\bar G}{dx}\right)_{x=x'-0}\right] =  -i\hbar,
                           \label{mf11a}\\
    \left[-\frac{\hbar^2}{2m} \frac{d^2}{dx^2}
    + V(x) - i\hbar s  \right] \bar G(x,x',s) =0, \qquad x\ne x'.
                                          \label{mf11b}
\end{gather}

We proceed now to indicate how we can derive the explicit expression
of $K(x,x',t)$ with the help of the Green function  $\bar G(x,x',s)$
of the corresponding problem satisfying (\ref{mf11a}) and
(\ref{mf11b}).

\section{Determination of the Green function\\ and the inverse
Laplace transform for the propagator}\label{section3}

Our interest is not to stop at equations (\ref{mf11a}), (\ref{mf11b}) for
$\bar G (x,x',s)$ but actually to get $K(x,x',t)$ for which we can
use the inverse Laplace transform~\cite{2} to get
            \begin{gather}
        K(x,x',t) = \frac{1}{2\pi i} \int^{c+i\infty}_{c-i\infty}
        \bar G (x,x',s) e^{st} ds,
                           \label{mf12}
        \end{gather}
where the integration takes place along a line in the complex plane
$s$ parallel to the imaginary axis and at a distance $c$ to it so
that all singularities of $\bar G(x,x',s)$ in the $s$ plane are on
the left of it.

To have a more transparent notation rather than the $s$ plane we
shall consider an energy variable $E$ proportional to it through the
relation
\[
    E= i \hbar s \qquad {\mbox{\rm or}}  \qquad s=-i (E/\hbar)
                           %\label{mf13}
\]
and def\/ine $G(x,x',E)$ by
\[
        -i G (x,x',E) \equiv \bar G (x,x', -iE/\hbar).
                           %\label{mf14}
\]

The energy Green function must be symmetric under interchange
of $x$ and $x'$, i.e.
            \begin{gather}
         G (x,x',E )=  G (x',x,E)
                           \label{mf15}
        \end{gather}
which combines with the two equations (\ref{mf11a}), (\ref{mf11b})
to give in this notation
\begin{gather}
    \left[\frac{dG}{dx}\right]_{x=x'+0} - \left[\frac{dG}{dx}\right]_{x=x'-0}
    = -\frac{2m}{\hbar},
                           \label{mf16a}\\
    \left[-\frac{\hbar^2}{2m}\frac{d^2}{dx^2} + V(x) - E\right]
     G(x,x',E) =0 \qquad {\rm for}\quad x\ne x'.
                                          \label{mf16b}
\end{gather}

Let us f\/irst consider the case when $x<x'$ and proceed to show that
the equations (\ref{mf15})--(\ref{mf16b}) determine
in a unique way the Green function of the problem. For this purpose
we introduce with the notation $u^\pm_E(x)$ two  linearly
independent solutions of the equation (\ref{mf16b})
\[
          \left[-\frac{\hbar^2}{2m}\frac{d^2}{dx^2} + V(x) - E\right]
          u^\pm_E (x) = 0.
                           %\label{mf17}
\]

From this equation we see that
\[
    u^-_E (x) \frac{d^2u^+_E (x)}{dx} - u^+_E \frac{d^2u^-_E (x)}
{dx} =  \frac{d}{dx} \left(u^-_E \frac{du^+_E}{dx} - u^+_E \frac{d
u^-_E}{dx}\right) =0.
                           %\label{mf18}
\]

Thus the Wronskian of the problem def\/ined by
            \begin{gather}
        W (E) = u^-_E (x) \frac{du^+_E}{dx}
        - u^+_E (x) \frac{du^-_E}{dx}
                           \label{mf19}
        \end{gather}
is independent of $x$.

As $G(x,x',E)$ satisf\/ies (\ref{mf16b}) we can write it for $x<x'$ as
            \begin{gather}
        G(x,x',E) = F(x', E) u^+_E (x),
                           \label{mf20}
        \end{gather}
choosing one of the two solutions of equation (\ref{mf16b}) and
$F(x',E)$ is as yet an undetermined function of $x'$, $E$.

We see from the symmetry of $G(x,x',E)$ that it must satisfy the
same equation (\ref{mf16b}) in $x'$ so that from (\ref{mf20}) we get
\[
         \left[-\frac{\hbar^2}{2m}\frac{d^2}{dx'^2} + V(x') - E\right] F(x',E)=0
                           %\label{mf21}
\]
and thus $F(x',E)$ must a be linear combination of the two
independent solutions $u^\pm_E(x)$, i.e.
\[
        F(x',E) = a_+ (E) u^+_E (x') + a_-(E) u^-_E (x')
                          % \label{mf22}
\]
and our Green function becomes
            \begin{gather}
    G(x,x',E) = \big[ a_+ (E) u^+_E (x') +  a_- (E) u^-_E (x') \big] u^+_E(x),
                           \label{mf23}
        \end{gather}
while for the other case, i.e.\ $x>x'$, the symmetry of the Green
function demands
                        \begin{gather}
    G(x,x',E) = \big[ a_+ (E) u^+_E (x) +  a_- (E) u^-_E (x) \big] u^+_E(x').
                           \label{mf23bis}
        \end{gather}
Replacing (\ref{mf23}) and (\ref{mf23bis}) in (\ref{mf16a}) we f\/ind
that the coef\/f\/icient $a_+(E)$ vanishes and $a_-(E)$ satisf\/ies
\begin{gather}
    a_-(E) W(E) = -\frac{2m}{\hbar}.
                           \label{mf25}
\end{gather}

Thus from (\ref{mf23}), (\ref{mf23bis}) and (\ref{mf25}) we get that
\begin{gather}
    G(x,x',E) = - \frac{2m}{\hbar} W^{-1} (E) \left\{ \begin{array}{c} u^-_E (x') u^+_E (x)
 \quad {\rm if } \  x < x', \vspace{1mm}\\ u^-_E (x) u^+_E (x') \quad {\rm if } \  x >x'.
 \end{array} \right.
                           \label{mf26}
\end{gather}

We thus have the explicit Green function of our problem once we can
obtain two independent solutions of the equations (\ref{mf16b}).

Once $G(x,x',E)$ has been determined, the propagator $K(x,x',t)$ is
given by the inverse Laplace transform (\ref{mf12}) which in terms
of the $E$ variable becomes
            \begin{gather}
            K(x,x',t) = \frac{1}{2\pi \hbar i} \int^{i\hbar c+
            \infty}_{i\hbar c-\infty} \exp (-iE t/\hbar) G(x,x',E) dE,
                           \label{mf27}
        \end{gather}
where now the integral takes place in the $E$ plane over a line
parallel to the real axis with  all the poles of $G(x,x',E)$ below
it.

We proceed to give some specif\/ic examples of application of our method.

\section[Specific examples]{Specif\/ic examples}\label{section4}

\subsection*{a) The free particle}

The potential $V(x)$ is taken as zero and so the equation
(\ref{mf16b}) becomes
\[
         \left[-\frac{\hbar^2}{2m}\frac{d^2}{dx^2}  - E\right]
         G(x,x',E) = 0.
                           %\label{mf28}
\]

We introduce the variable $k$ through the def\/inition
\[
        E= \frac{\hbar^2k^2}{2m} , \qquad dE = \frac{\hbar^2k}{m} dk
                           %\label{mf29}
\]
and thus the $u^\pm_E(x)$ for this problem satisfy the equation
\[
        \left[ \frac{d^2}{dx} + k^2\right] u^\pm_E (x) = 0, \qquad u^\pm_E (x) = \exp (\pm i k x)
                                  %\label{mf30}
\]
with the Wronskian (\ref{mf19}) given by
\[
        W(E) = 2 i k.
                          % \label{mf31}
\]

Thus from the two cases of (\ref{mf26}) our function $G(x,x',E)$ is
written compactly as
            \begin{gather}
        G(x,x'E) = \frac{im}{\hbar k} \exp  [ik |x-x'|].
                           \label{mf32}
        \end{gather}
The propagator $K(x,x',t)$ is given by (\ref{mf27}) in terms of
$G(x,x',E)$ and substituting (\ref{mf32}) in it and writing in terms
of $k$ we get
            \begin{gather}
        K(x,x',t) = \frac{1}{2\pi} \int^\infty_{-\infty} \exp [i k |x-x'| -
        i (\hbar k^2/2m) t] dk,
                           \label{mf33}
        \end{gather}
where, as $G(x,x',E)$ has no singularities, the energy can be
integrated over the real line \mbox{$-\infty \leq E \leq \infty$} while $k$
has double the range of $E$.

The integral (\ref{mf33}) can be determined by completing the square
and we get
            \begin{gather}
            K(x,x',t) = \sqrt{\frac{m}{2\pi i \hbar t} } \exp \left[
                \frac{im(x-x')^2}{2\hbar t}\right]
                           \label{mf34}
        \end{gather}
which has been derived also by many other methods.

\subsection*{b) The case of the delta potential}

We wish now to discuss the ef\/fect on the Feynman propagator of a
potential
\[
           V(x) = Q(x) + b\delta (x),
                           %\label{mf37}
\]
where $Q(x)$ in a continuous function of $x$ and we assume $b>0$
to avoid bound states of the $\delta$ potential.

The equation for $u^\pm_E$ becomes now
\[
           [H + b\delta (x) - E] u^\pm_E (x) = 0,
                           %\label{mf38}
\]
where
\[
        H=- \frac{\hbar^2}{2m}  \frac{\partial^2}{\partial x^2}
          + Q (x).
                           %\label{mf39}
\]

The Green function $G(x,x', E)$ satisf\/ies the equations
(\ref{mf16a}),  (\ref{mf16b}) which can be written as the single
equation
                  \begin{gather}
        [H + b \delta (x) -E] G (x,x',E) = -i \hbar \delta (x-x')
                           \label{mf40}
        \end{gather}
and (\ref{mf16a}) holds if we integrate (\ref{mf40}) with respect to
the variable $x$  in the integral $x' - \epsilon \leq x \leq x'+
\epsilon$ in the limit $\epsilon\to 0$, and the one corresponding to
(\ref{mf16b}) holds when $x\ne x'$.

For $x \not= 0$ we have (\ref{mf40}) with no delta potential and
therefore $G$ can be written as
             \begin{gather}
     G(x,x',E) = G_Q(x,x',E) + F(x,x',E), \qquad x \not= 0,
                                          \label{mf41}
\end{gather}
where $G_Q(x,x',E)$ is the Green function satisfying
    \begin{gather}
[H - E ]
     G_Q(x,x',E) = -i \hbar \delta(x-x')
                                          \label{mf42}
        \end{gather}
while $F(x,x',E)$ is a solution of the corresponding homogeneous
equation, i.e.
\[
[H - E ]
     F(x,x',E) = 0, \qquad x \not= x'.
                                         % \label{mf43}
\]
and the form of $F(x,x',E)$ is to be determined. The continuity of
$G(x,x',E)$ at $x=0$ allows to write (\ref{mf41}) for all values of
$x, x'$ and with this in mind we can replace (\ref{mf41}) in
(\ref{mf40}) to obtain
\begin{gather}
[H - E - b \delta(x) ]
[ G_Q(x,x',E) + F(x,x',E) ] =- i \hbar \delta(x-x') \nonumber \\
\qquad{}= -i \hbar \delta(x-x') -b \delta(x)G_Q(0,x',E) +
[H - E - b \delta(x) ]F(x,x',E),
                                          \label{mf44}
\end{gather}
where in the second line we have used (\ref{mf42}). The two lines in
(\ref{mf44}) imply
\[
[H - E - b \delta(x) ]F(x,x',E) = b \delta(x)G_Q(0,x',E)
                                          %\label{mf45}
\]
which is a version of (\ref{mf40}) but with a source term. Therefore
the solution can be readily given as
\[
    F(x,x',E) = -\frac{i}{\hbar}\int^{\infty}_{-\infty} dx'' G(x,x'',E)
   ( b \delta(x'')G_Q(0,x',E)).
                                          %\label{mf46}
\]
Performing the integral in the last expression and using
(\ref{mf41}) for $G$, we have
            \begin{gather}
    F(x,x',E) = \gamma \left[ G_Q(x,0,E) + F(x,0,E) \right] G_Q(0,x',E),
                                          \label{mf47}
        \end{gather}
where $\gamma \equiv -ib/\hbar $. To determine $F(x,0,E)$ in the RHS
of (\ref{mf47}) we set $x'=0$ and solve for $F$, obtaining
\begin{gather}
    F(x,0,E) = \frac{\gamma G_Q(x,0,E) G_Q(0,0,E)}{1-\gamma G_Q(0,0,E)}.
                                          \label{mf48}
\end{gather}
Finally, (\ref{mf48}) can be replaced back in (\ref{mf47}) to get
\begin{gather}
    F(x,x',E) = \frac{\gamma G_Q(x,0,E) G_Q(0,x',E)}{1-\gamma G_Q(0,0,E)}.
                                          \label{mf49}
\end{gather}
With this, $G(x,x',E)$ is given now in terms of $G_Q(x,x',E)$ and if
we make $Q=0$ we can apply it to the case of the free particle. The
Green function $G_0(x,x',E)$ is given by (\ref{mf32}) and the Green
function of our problem becomes
\begin{gather*}
   G(x,x',E) = \frac{im}{\hbar k} \exp  [ik |x-x'|] -
   \frac{m^2 b}{2\hbar^4}\frac{\exp  [ik (|x|+|x'|)]}{k \left(k +i\frac{mb}{\hbar^2}
   \right)}\nonumber\\
\phantom{G(x,x',E)}{} = \frac{im}{\hbar k} \exp  [ik |x-x'|] -
   \frac{ i m }{2\hbar^2}\frac{\exp  [ik (|x|+|x'|)]}{ k } +
    \frac{ i m }{2\hbar^2}\frac{\exp  [ik (|x|+|x'|)]}{k+i\frac{mb}{\hbar^2}}.
                           %\label{mf50}
\end{gather*}
This is the same result that appears in Grosche and Steiner~\cite[(6.12.4), p.~328]{1}
and  accounts explicitly for the four
possible cases $\pm x, \pm x'
> 0$. The inverse Laplace transform of this expression can be easily
evaluated since it contains integrals of the form
\[
        \int^{\infty}_{-\infty} dk k^{-n}
\exp \left( -C k^2 + D k \right), %\label{mf51}
\]
where $C$ and $D$  are independent of $k$ and the integrals are
either gaussians or error functions when $n=0$ or $n=1$
respectively. The propagator is then
\begin{gather*}
                K(x,x',t)= K_0(x,x',t)+
        \frac{mb}{2\hbar^2} \exp \left( -\frac{mb}{ \hbar^2 }\left( |x|+|x'|+
        \frac{imb^2 t}{2 \hbar^3}\right) \right) \nonumber \\
\phantom{K(x,x',t)=}{} \times { \rm erfc  }
        \left\{ \sqrt{\frac{m}{2i \hbar t}} \left( |x|+|x'|-\frac{ibt}{ \hbar }
                    \right) \right\} %\label{mf52}
\end{gather*}
with $K_0(x,x',t)$ as in (\ref{mf34}). This result coincides with
the one that Grosche and Steiner~\cite[(6.12.2), p.~327]{1} obtained by the Path
Integral Methods.

Note also that if in $G$ of (\ref{mf41}) we replace the $F(x,x',E)$
by (\ref{mf49}) we also get the result of Grosche and Steiner~\cite[(6.12.1), p.~327]{1}.

\subsection*{c) The harmonic oscillator}

The potential $V(x)$ is proportional to $x^2$ and thus $u^\pm_E(x)$
satisf\/ies the equation
            \begin{gather}
        \left[ - \frac{\hbar^2}{2m} \frac{d^2}{dx^2}
        + \frac12 m \omega^2 x^2 - E\right]
        u^\pm_E (x) = 0,
                           \label{mf35a}
        \end{gather}
where $\omega$ is the frequency of the oscillator.

We introduce the variables
            \begin{gather}
    z = \sqrt{\frac{2m\omega}{\hbar} } x , \qquad p = \frac{E}{\hbar\omega} - \frac12
                           \label{mf36a}
        \end{gather}
in terms of which the equation (\ref{mf35a}) takes the form
            \begin{gather}
    \left[ \frac{d^2}{dz^2} - \frac{z^2}{4} +
    p + \frac12 \right] u^\pm_E(x) =0.
                           \label{mf37a}
        \end{gather}
Two independent solutions of (\ref{mf37a}) are given by parabolic
cylinder functions~\cite{3}, i.e.
\[
    u^\pm_E (x) = D_p(\pm z).
                           %\label{mf38a}
\]

The Wronskians of these functions, where the derivative is taken
with respect to the $x$ rather than the $z$ variable, is from
(\ref{mf19}) and (\ref{mf36a}) given by
\begin{gather*}
            W(E) = \sqrt{\frac{2m\omega}{\hbar} }
                \left\{ D_p (-z) \left[ \frac{dD_p(z)}{dz}\right]
                - D_p (z) \frac{dD_p(-z)}{dz}\right\}\\
                %\label{mf39a}\\
\phantom{W(E)}{} = \sqrt{\frac{2m\omega}{\hbar}}
                \bigg\{ D_p (-z) \left[ -D_{p+1}(z) + \frac12 zD_p(z)\right]
                + D_p(z) \left[ - D_{p+1} (-z) - \frac12 z
                D_p(-z)\right] \bigg\},\nonumber
\end{gather*}
where we made use \cite[(9.247-3), p.~1066]{3}
together with the fact that
\[
    [dD_p (-z)/dz] = - [d D_p(-z)/d(-z)].
                           %\label{mf40a}
\]

The Wronskian is independent on $z$ so we may take any value of the
latter and we choose $z=0$ to get
\[
        W (E) = - \sqrt{\frac{2m\omega}{\hbar} } 2 D_p (0) D_{p+1} (0).
                           %\label{mf41a}
\]

We note from \cite[(9.240), p.~1064]{3} that we can
write $D_p(z)$ in terms of the degenerate hypergeometric function
$\Phi$ and, in particular
\[
    D_p (0) = 2^{p/2} \frac{\sqrt{\pi}}{\Gamma \big(\frac{1-p}{2}\big)}
    \Phi \left(- \frac{p}{2}, \frac12, 0\right)
                               %\label{mf42a}
\]
while from \cite[(9.210), p.~1058]{3} the $\Phi (-\frac{p}{2}, \frac12, 0) =1$ so
that  f\/inally
            \begin{gather}
        W(E) = - \sqrt{\frac{2m\omega}{\hbar}} \frac{2^{p+1} \pi}{\Gamma
        \big(\frac{1-p}{2}\big) \Gamma \big(- \frac{p}{2}\big)}
        = - \sqrt{ \frac{2m\omega}{\hbar}} \frac{\sqrt{\pi}}{\Gamma(-p)},
                           \label{mf43a}
        \end{gather}
where for the last expression in (\ref{mf43a}) we made use of  the
doubling formula for the $\Gamma$ function given in \cite[(8.335), p.~938]{3}.

From the general relation (\ref{mf26}) we then obtain that the Green
function of the oscillator is given by
            \begin{gather}
        G(x,x',E) = \sqrt{\frac{2m}{\pi \hbar \omega}}
        \Gamma(-p) D_p (z) D_p (-z')
                               \label{mf44a}
        \end{gather}
with $p$, $z$ given by (\ref{mf36a}), $z<z'$ and $z'$ has the same
def\/inition as $z$ but $x$ replaced by $x'$. When we consider the
case $z>z'$ we have an expression which is similar to (\ref{mf44a})
but interchanging $z$ and $z'$. The complete formula can be written
in a compact way introducing the variables
\[
            x_> = \max \{ x,x' \} , \qquad x_< = \min \{ x,x' \}
                           %\label{mf45a}
\]
and thus we have
            \begin{gather}
            G(x,x',E) = \sqrt{\frac{2m}{\pi \hbar  \omega}} \Gamma
            \left(\frac12 - \frac{E}{\hbar\omega}\right)
            D_{\frac{E}{\hbar\omega} - \frac12} \left( \sqrt{\frac{2m\omega}{\hbar}}
            x_> \right) D_{\frac{E}{\hbar\omega} -\frac12}
            \left(-\sqrt{\frac{2m\omega}{\hbar}} x_<\right).
                          \label{mf46a}
      \end{gather}

The expression (\ref{mf46a}) is identical to \cite[(6.2.37), p.~179]{1}
except for a factor of $\sqrt2$.

We want though to obtain $K (x,x',t)$ of (\ref{mf27}), but for this
we can analyze the pole structure of (\ref{mf46a}) in order to
evaluate the inverse Laplace transform of $G(x,x',E)$ by means of
the residue theorem. The result of this procedure is the series
known as the spectral decomposition of $K (x,x',t)$. This series can
be evaluated by using an identity  of Hermite polynomials in~\cite{4} and references cited therein. We carry out the analysis in
the Appendix and get the f\/inal result
\[
    K(x,x',t) = \left( \frac{m\omega}{2\pi i \hbar \sin\omega
    t}\right)^{1/2}
    \exp \left\{ \frac{im\omega}{2\hbar \sin\omega t} \bigg[ (x'^2 +x^2)
    \cos \omega t - 2 xx'\bigg]\right\}
                           %\label{mf48a}
\]
which coincides with expression in \cite[p.~160]{1}.

\section{Conclusion}\label{section5}

We will state here the full steps to get the Feynman propagator of a
non-relativistic single particle one dimensional problem.

Our Hamiltonian is
\[
    H= \left[ - \frac{\hbar^2}{2m} \frac{\partial^2}{\partial x^2} +
    V(x)\right]
                           %\label{mf74}
\]
with an arbitrary potential $V(x)$.

We assume that two independent eigen-functions of energy $E$ can be
found and denoted by~$u^\pm_E (x)$.
%\[
%           u^\pm_E (x).
%                           %\label{mf75}
%\]

We need then to determine the Wronskian
\[
            W(E)= u^-_E (x) \frac{du^+_E (x)}{dx} - u^+_E (x) \frac{d
            u^-(x)}{dx}
                           %\label{mf76}
\]
whose value, as we know, is independent on $x$.

Following the analysis of Section~\ref{section3} we can determine the energy
Green function
\[
            G(x,x',E)= - \frac{2m}{\hbar} W(E)^{-1}
            \left\{
              \begin{array}{l}
                u^-_E (x') u^+_E (x) \quad  \hbox{if}\  x<x', \vspace{1mm}\\
                 u^-_E (x) u^+_E (x') \quad  \hbox{if} \  x>x'.
              \end{array}
            \right.
                          % \label{mf77}
\]

Using the Laplace transform the Feynman propagator becomes
                \begin{gather}
    K(x,x',t) = \frac{1}{2\pi \hbar i} \int^{i\hbar
    c+\infty}_{i\hbar c-\infty} \exp (-i E t/\hbar) G(x,x',E) dE.
                           \label{mf78}
        \end{gather}

Once we get $u^\pm_E (x)$ the only troublesome part of our
calculation is the integral (\ref{mf78}) as we already saw in the
discussion of the examples in Section~\ref{section4}.

For time dependent Hamiltonians probably other techniques should be
used but we have not developed them yet.

\appendix

\section{Determination of the propagator for the harmonic
oscillator}\label{sectionA}

We shall proceed to express the Green function $G(x,x', E)$ in terms
of its energy poles and residues and then apply the Laplace
transform to it to get the propagator.

\begin{figure}[t]
\centerline{\includegraphics[width=3in]{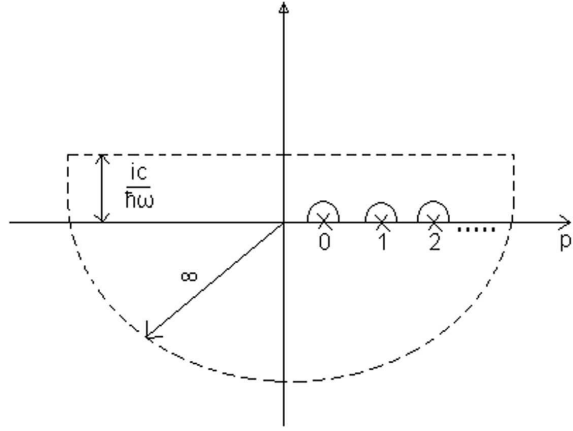}}
\caption{Integration contour for the determination of the propagator.\label{fig1}}
\end{figure}

To achieve our objective we start with the complex variable integral
                          \begin{gather}
            \frac{\omega}{2\pi i}   \int_C
            e^{-i\hbar\omega\big(p+\frac 12\big)t}A \Gamma (-p) D_p(z)
            D_p(-z) dP,
                           \label{amf70}
        \end{gather}
where
                          \begin{gather}
            p=\frac{E}{\hbar\omega} - \frac12, \qquad
            z=\sqrt{\frac{2m\omega}{\hbar}} x, \qquad
            z'= \sqrt{\frac{2m\omega}{\hbar}} x', \qquad
            A= \sqrt{\frac{2m}{\pi \hbar  \omega}} ,
                           \label{amf71}
        \end{gather}
and the contour $C$ of integration is depicted in
Fig.~\ref{fig1}.

The function $\Gamma$ is gamma while the parabolic cylinder is
function $D_p(z)$. If we substitute $p$, $z$, $z'$ of (\ref{amf71}) in the
integration (\ref{amf70}) and replace the contour in Fig.~\ref{fig1}
just by its upper line we obtain the integral
\begin{gather*}
              K (x,x',t')= \frac{1}{2\pi \hbar i}
              \int^{i\hbar c+\infty}_{i\hbar  c -\infty} \exp(-i E
              t/\hbar ) \sqrt{\frac{2m}{\hbar^3 \pi \omega} }
              \Gamma \left(\frac12 -\frac{E}{\hbar\omega}\right) D_{\frac{E}{\hbar \omega} -
              \frac 12}(z) D_{\frac{E}{\hbar \omega} -
              \frac 12}(-z') dE.
                          % \label{amf73}
\end{gather*}

We proceed now to evaluate integral (\ref{amf70}) in terms of energy
poles and residues of its integrand.

The parabolic cylinder functions $D_p(z)$, $D_p(-z')$ are analytic in
the full $p$ plane as shown in \cite[(2), p.~1066]{3},
while $\Gamma (-p)$ has poles only on the real $p$ axis with values
                      \begin{gather}
           p= n= 0,1,2, \dots.
                           \label{amf74}
                   \end{gather}

Note also that the residues  of $\Gamma(-p)$ at the poles given in
(\ref{amf74}) are{\samepage
% $\frac{(-)^n}{n!}$
\[
                      {\rm Res}\, \Gamma(-p) = \frac{(-1)^n}{n!}
                                   %\label{amf75}
\]
as shown at~\cite[p.~933]{3}.}

        Furthermore the integral
\[
         \frac{1}{2\pi i} \int^\infty_{-\infty} \frac{D_p(z)
         D_p(-z')}{p-n} dp = D_n (z) D_n(-z')
                                   %\label{amf76}
\]
as $D_p(z)$ is analytic in the full $p$ plane.

The lower circle in the contour of Fig.~\ref{fig1} does not contribute to
the integral because the term $\exp (-i\omega tp)$ vanishes when
$p\to \infty$ with a negative imaginary part. As the poles marked
with across in Fig.~\ref{fig1} are only on the real $p$ axis we can
eliminate the region between the upper line and the real axis and
all that remains is the residue of $\Gamma (-p)$ in the integrand
(\ref{amf70}) so $K(x,x',t)$ takes the form
\[
        K(x,x',t) = \omega \sqrt{\frac{2m}{\pi \hbar 
        \omega}} \sum^\infty_{n=0} \frac{(-1)^n}{n!} D_n \left(
        \sqrt{\frac{2m\omega}{\hbar}} x\right) D_n
        \left( -  \sqrt{\frac{2m\omega}{\hbar}} x'\right) e^{-i\omega
        \big(n+\frac 12\big)t}.
        % \label{amf77}
\]
        The parabolic cylinder of index $n$ can be put in terms of
        an Hermite polynomial
\[
         D_n  (z) = 2^{-\frac{n}{2}} \; e^{-\frac{z^2}{4}} H_n \left(
         \frac{z}{\sqrt2}\right),
         %\label{amf77a}
\]
so the propagator can also be written as
    \begin{gather*}
 K(x,x',t) = \sqrt{\frac{2m\omega}{\hbar\pi} }
        \sum^\infty_{n=0} \frac{e^{-i\omega
        (n+1/2)t}}{n!2^n}\nonumber\\
  \phantom{K(x,x',t) =}{}\times\exp\left\{ -\frac{\omega m}{2\hbar} (x^2 + x'^2)\right\}
      H_n \left( \sqrt{\frac{m\omega}{\hbar}} x\right)
      H_n \left( -\sqrt{\frac{m\omega}{\hbar}} x'\right).
                           %\label{amf78}
\end{gather*}

Using the relation \cite{4}
\begin{gather*}
 \sum^\infty_{n=0} \frac{H_n(z) H_n(z')}{n!} \xi^n
      = \frac{1}{\sqrt{1-4\xi^2}}
      \exp \left\{
      \frac{2\xi (2\xi (z^2 + z'^2)-2zz')}{4\xi^2 -1} \right\},
                           %\label{amf79}
\end{gather*}
we obtain
\begin{gather}
            K(x,x'; t)   =  \sqrt{\frac{2m\omega}{\hbar \pi}
            } e^{-\frac{i\omega t}{2}} \exp \left\{
            -\frac{m\omega}{2\hbar} (x^2 + x'^2)\right\}
            \frac{1}{\sqrt{1-e^{-i2\omega t}}}
            \nonumber\\
\phantom{K(x,x'; t)   =}{} \times \exp \left\{ \frac{m\omega}{\hbar}
            \frac{e^{-i\omega t}(e^{-i\omega t}(x^2 + x'^2) -2
            xx')}{e^{-2i\omega t}-1} \right\}
            \nonumber\\
\phantom{K(x,x'; t)  }{}= \sqrt{\frac{2m\omega}{\pi \hbar}}
            \frac{1}{\sqrt{e^{i\omega t} - e^{-i\omega t}}} \exp
            \left\{ \frac{m\omega}{\hbar} \left[ (x^2 + x'^2) \left(
            \frac{e^{-2i\omega t}}{e^{-2i\omega t}-1} -\frac12\right)- \frac{2
            e^{-i \omega t} xx'}{e^{-2i\omega t}-1} \right]
            \right\}
            \nonumber\\
\phantom{K(x,x'; t)  }{}= \sqrt{\frac{m\omega}{2i\pi \hbar \sin \omega t}} \exp \left\{
    \frac{im\omega}{2\hbar \sin (\omega t)} \big[ \cos (\omega t)
    (x^2 + x'^2) - 2 xx'\big]\right\}.
                           \label{amf80}
\end{gather}

As equation (\ref{amf80}) is identical to (\ref{mf78}) and this also
applies to the other corresponding equations discussed in the
examples, we see that our alternative approach is equivalent to that
of Feynman.

Other methods for deriving the propagator for the case of the
harmonic oscillator have been proposed recently~\cite{5,6}.

\subsection*{Acknowledgements}

One of the authors (A. del Campo) would like to express his thanks
for the hospitality of the Instituto de F\'isica and the support
both from the Instituto de F\'isica and of CONACYT (Project No.~40527F) for the time he spent in Mexico. This author
would also like to thank the Basque Government (BFI04.479) for
f\/inancial support. E.~Sadurn\'i is grateful to CONACYT and its
support through Beca-Cr\'edito 171839. M.~Moshinsky is grateful to
his secretary Fanny Arenas for the capture of this manuscript and
the 300 she has done previously.

\pdfbookmark[1]{References}{ref}
\LastPageEnding

\end{document}